\documentclass[5p,times,amsmath,amssymb]{elsarticle}

\pdfoutput=1
\usepackage{lineno,hyperref}
\usepackage{color}
\usepackage{siunitx}
\modulolinenumbers[5]

\journal{Physics Letters B}

\begin{document}

\begin{frontmatter}

\title{Search for chameleons with CAST}

\author[Patras]{V.~Anastassopoulos}
\author[dogus]{M.~Arik\fnref{arik}}   \fntext[arik]{Pr. addr.: Bogazici University, Istanbul, Turkey.}
\author[Saclay]{S.~Aune}
\author[CERN]{K.~Barth}
\author[Moscow]{A.~Belov}
\author[MPE]{H.~Br{\"a}uninger}
\author[INFN]{G.~Cantatore}
\author[Zgz]{ J.~M.~Carmona}
\author[dogus]{S.~A.~Cetin}
\author[DTU]{F.~Christensen}
\author[Fermi]{J.~I.~Collar}
\author[Zgz]{T.~Dafni}
\author[CERN]{M.~Davenport\corref{mycorrespondingauthor}}
\ead{Martyn.Davenport@cern.ch}
\author[Bonn]{K.~Desch}

\author[Moscow]{A.~Dermenev}

\author[AUTH]{C.~Eleftheriadis}
\author[Demokritos]{G.~Fanourakis}
\author[Saclay]{E.~Ferrer-Ribas}
\author[MPE]{ P.~Friedrich}
\author[Saclay]{J.~Gal\'{a}n}
\author[Zgz]{J.~A.~Garc\' ia}
\author[Patras]{A.~Gardikiotis}
\author[Zgz]{J.~G.~Garza}
\author[NTUA]{E.~N.~Gazis}
\author[Demokritos]{T.~Geralis}
\author[Saclay]{I.~Giomataris}
\author[CU]{C.~Hailey}
\author[CERN]{F.~Haug}
\author[Columbia]{M.~D.~Hasinoff}
\author[Darmstad]{D. H. H. Hoffmann}
\author[Zgz]{F.~J.~Iguaz}
\author[Zgz]{I.~G.~Irastorza}
\author[Goethe]{J.~Jacoby}
\author[DTU]{A.~Jakobsen}
\author[Zagreb]{K.~Jakov\v{c}i\'{c}}
\author[Bonn]{J.~Kaminski}
\author[Rijeka,INFN]{M.~Karuza}
\author[dogus]{M.~Kavuk\fnref{kavuk}}   \fntext[kavuk]{Pr. addr.: Bogazici University, Istanbul, Turkey.}
\author[Zagreb]{M.~Kr\v{c}mar}
\author[Bonn]{C.~Krieger}
\author[CERN]{A.~Kr\"{u}ger\fnref{kruger}}   \fntext[kruger]{Pr. addr.: Hochschule Karlsruhe Technik und Wirtschaft, Univ. of Applied Sciences, Moltkestr. 30, 76133 Karlsruhe, Germany.}
\author[Zagreb]{B.~Laki\'{c}}
\author[CERN]{J.~M.~Laurent}
\author[AUTH]{A.~Liolios}
\author[Zagreb]{A.~Ljubi\v{c}i\'{c}}
\author[Zgz]{G.~Luz\'on}
\author[Darmstad]{S.~Neff}
\author[Zgz,CERN]{I.~Ortega}
\author[Saclay]{T.~Papaevangelou}
\author[Livermore]{M.~J.~Pivovaroff}
\author[MPIphys]{G.~Raffelt}
\author[Darmstad]{H.~Riege}
\author[Darmstad]{M.~Rosu}
\author[Livermore]{J.~Ruz}
\author[AUTH]{I.~Savvidis}
\author[MPIsonn]{S.~K.~Solanki\fnref{solanki}}   \fntext[solanki]{Sec. affiliation: School of Space Research, Kyung Hee University, Yongin, Republic of Korea.}
\author[CERN,AUTH]{T. Vafeiadis\corref{mycorrespondingauthor}}
\ead{Theodoros.Vafeiadis@cern.ch}
\author[Zgz]{J.~A.~Villar}
\author[Livermore]{J.~K.~Vogel}
\author[dogus]{ S.~C.~Yildiz\fnref{yildiz}}   \fntext[yildiz]{Pr. addr.: Dep. of Physics and Astronomy, University of California Irvine, Irvine, CA 92697, USA.}
\author[CERN,Patras]{K.~Zioutas}

\author[]{(CAST~Collaboration)}
\author[]{and~P.~Brax\fnref{brax}} \fntext[brax]{Affiliation: Institut de Physique Th\'eorique, CEA, IPhT, CNRS, URA 2306, F-91191Gif/Yvette Cedex, France.}

\author[]{I.~Lavrentyev\fnref{ivan}}   \fntext[ivan]{Affiliation: Boston University, Boston, MA 02215, USA.}
\author[]{A.~Upadhye\fnref{upadhye}}   \fntext[upadhye]{Affiliation: Physics Dep., University of Wisconsin-Madison, 1150 University Avenue, Madison, WI 53706, USA.}

\address[Patras]{Physics Department, University of Patras, Patras, Greece}
\address[dogus]{Dogus University, Istanbul, Turkey}
\address[Saclay]{IRFU, Centre d’ Etudes Nucl\'{e}aires de Saclay (CEA-Saclay), Gif-sur-Yvette, France}
\address[CERN]{European Organization for Nuclear Research (CERN), Gen\`{e}ve, Switzerland}
\address[Moscow]{Institute for Nuclear Research (INR), Russian Academy of Sciences, Moscow, Russia}
\address[MPE]{Max-Planck-Institut f\'{u}r Extraterrestrische Physik, Garching, Germany}
\address[INFN]{Istituto Nazionale di Fisica Nucleare (INFN), Sezione di Trieste and Universit\`a di Trieste, Trieste, Italy}
\address[Zgz]{Instituto de F\'{i}sica Nuclear y Altas Energ\'{i}as, Universidad de Zaragoza, Zaragoza, Spain}
\address[DTU]{Danish Technical University-Space (DTU), Copenhagen, Denmark}
\address[Fermi]{Enrico Fermi Institute and KICP, University of Chicago, Chicago, IL, USA}
\address[Bonn]{Physikalisches Institut, Universit{\"a}t of Bonn, 53115 Bonn, Germany}
\address[AUTH]{Aristotle University of Thessaloniki, Thessaloniki, Greece}
\address[Demokritos]{National Center for Scientific Research ``Demokritos'', Athens, Greece}
\address[NTUA]{National Technical University of Athens, Athens, Greece}
\address[CU]{Columbia University (CU), New York, United States of America}
\address[Columbia]{Department of Physics and Astronomy, University of British Columbia, Vancouver, Canada}
\address[Darmstad]{Technische Universit{\"a}t Darmstadt, IKP, Darmstadt, Germany}
\address[Goethe]{J. W. Goethe-Universit{\"a}t, Institut f\'{u}r Angewandte Physik, Frankfurt am Main, Germany}
\address[Zagreb]{Rudjer Bo\v{s}kovi\'{c} Institute, Zagreb, Croatia}
\address[Rijeka]{Physics Department and Center for Micro and Nano Sciences and Technologies, University of Rijeka, Croatia}
\address[Livermore]{Lawrence Livermore National Laboratory, Livermore, CA 94550, USA}
\address[MPIphys]{Max-Planck-Institut f\'{u}r Physik (Werner-Heisenberg-Institut), M{\"u}nchen, Germany}
\address[MPIsonn]{Max-Planck-Institut f\"{u}r Sonnensystemforschung, G\"{o}ttingen, Germany}

\cortext[mycorrespondingauthor]{Corresponding author.}

\begin{abstract}

 In this work we present a search
for (solar) chameleons with the CERN Axion Solar Telescope (CAST).
 This novel experimental technique, in the field of dark energy research, exploits both
the chameleon coupling to matter ($\beta_{\rm m}$) and to photons ($\beta_{\gamma}$) via the Primakoff effect.
By reducing the X-ray detection energy threshold used for axions from 1$\,$keV to 400$\,$eV
CAST became sensitive to the converted solar chameleon spectrum 
which peaks around 600$\,$eV. Even though we have not observed any
excess above background, we can provide a 95\% C.L. limit for
the coupling strength of chameleons to photons of $\beta_{\gamma}\!\lesssim\!10^{11}$  for $1<\beta_{\rm m}<10^6$.

\end{abstract}

\begin{keyword}
\texttt chameleon \sep CAST \sep SDD \sep X-ray \sep tachocline \sep dark energy
\MSC[2015] 00-01\sep  99-00
\end{keyword}

\end{frontmatter}


\section{Introduction}

The dark sector of cosmology represents a big challenge in fundamental  physics. In particular, dark energy~\cite{Cop06,Cli12},
which is responsible for the accelerated expansion of the Universe, could be due to the existence of  a scalar field like the postulated chameleon~\cite{Kho04,Kho04_2,Bra04} (for a comprehensive theoretical treatment we refer to~\cite{Joy14}).
Although a high energy description of chameleons  derived from an ultraviolet completion such as string theory is still missing,
this type of low energy model is suggestive enough to justify novel investigations like the one presented in this work.

Chameleons can be created in the sun via the Primakoff effect. Like axions, creation could occur in the nuclear coulomb field of the plasma at the solar core, but such a calculation does not exist as yet, though it would be of interest. Additionally they can be created in regions of strong transverse magnetic fields in the solar interior. The tachocline, a region inside the Sun at a distance of around $0.7\,R_\odot$ from the centre, is widely believed to be the source of intense magnetic fields. At present only the characteristics of chameleon creation at the tachocline have been studied in detail, together with their propagation in the sun and journey to the helioscope ~\cite{Bra10, Bra12}.

Chameleons have non-linear self-interactions and interactions with matter which give them an ``effective mass''  dependent on the ambient mass (energy) density. 
The outer solar magnetic fields can transform chameleons to soft X-rays. The same could also happen with the integrated transverse magnetic field all the way from the Sun to the Earth, because the effective mass of chameleons decreases with lower and lower density in the free space between the Sun and the Earth’s atmosphere.
Taking into account the limit of $\beta_{\gamma}$ which saturates the solar luminosity, the transformation probability in any of the aforementioned magnetic fields is negligibly small and does not affect the expected flux arriving in CAST. Traversing the Earth$'$s atmosphere makes no difference in the intensity of the relatively energetic solar chameleons we are considering here.

They would have a very small effective mass in outer space or in the evacuated magnet cold bores of CAST~\cite{Zio99} but a large effective mass inside the detector material of most terrestrial dark matter experiments. Their corresponding energies generally exceed the chameleon effective mass inside matter and thus they traverse materials with hardly any interaction, making detection difficult.

Chameleon dark energy is an effective field theory (EFT) with a cutoff around the dark energy scale, 2$\,$meV, above which rigorous predictions cannot be made without a UV completion.  A single particle with a 3-momentum much higher than the cutoff is perfectly consistent with an EFT treatment, since the 3-momentum is not Lorentz invariant. However, we cannot rigorously quantify two-particle interactions with center-of-mass energies far above the cutoff. In one such process, fragmentation, two chameleons interact to form a greater number of lower-energy chameleons. The treatment of chameleon fragmentation as a coherent, semiclassical process in~\cite{Bra14} is inapplicable to the Sun.

In order for fragmentation in the Sun to be calculable within the EFT context, the cut-off of the higher order interactions leading to fragmentation would have to be increased towards a few keV from its nominal value of 2 meV. The way of doing this is not known, but changing the cut-off scale of chameleon models is a priori independent from the matter density dependence of both the vev of the chameleon and its mass. 
Hence one may envisage that models with a higher cut-off and the same dependence of the vev and the mass on the environment could be constructed. Since a UV completion is beyond the scope of this paper~\cite{Hin11} our analysis is made on the assumption that fragmentation at the tachocline  is negligible.

To investigate the existence of exotica like chameleons we present
the first results with a new experimental technique~\cite{Bra10}, by transforming
 an axion helioscope to a chameleonic one.

Cha\-meleons, like axions, can be detected
by the inverse Primakoff effect inside a transverse magnetic field, with their
conversion efficiency being optimised in vacuum.
Their expected spectrum originates from the photon thermal
spectrum and is modified by the creation probability in such an environment
which is proportional to the square of the photon energy ($\omega^2$) times a
factor of $\omega^{-1/2}$ which derives from the
fact that the photons perform a random walk in the Sun. All in all, 
this shifts the peak of the spectrum of produced chameleons from the photon
temperature in the tachocline at around 200$\,$eV to a much larger value
of 600$\,$eV. It is interesting to note that below $\sim$1$\,$keV 
the conversion probability from chameleon to photon via the Primakoff
effect is quasi-constant~\cite{Bra12}.

In this paper we discuss the upper limit on the chameleon to photon coupling strength ($\beta_\gamma$) for a wide range of their coupling to matter. Our result on $\beta_\gamma$
is comparable with the one obtained by the GammeV-CHASE (hereafter CHASE) experiments  in a laser cavity~\cite{Ste10, Upa12_2}. 
We explore uncertainties in the tachocline magnetic field, the precise radius and width of this region and the fraction of solar luminosity 
emited as chameleons. We also consider higher powers of the chameleon potential and show that our 
limit, $\beta_\gamma\!\lesssim\!10^{11}$, stands almost independent of the type of inverse power law potential used.

\section{The experiment}

The detection of solar chameleons can be performed with an axion helioscope
like CAST via the inverse Primakoff effect. The relevant chameleon-to-photon
coupling strength $\left(\beta_{\gamma}\right)$ replaces
the axion-to-photon coupling constant $\left(g_{a\gamma}\right)$.
 The expected X-ray spectrum peaks at 600$\,$eV, whereas the X-rays from axions from 
 the solar core are expected to appear with energies in the multiple
keV range. Therefore to study solar chameleons in CAST the cold bores should
be in vacuum, whilst the detector should ideally be sensitive to the
$150-1500\,$eV energy range.

After running with vacuum in the magnet bores in 2003 and 2004~\cite{Zio05, And07},
and with helium-filled cold bores between 2005 and 2012~\cite{Ari09, Ari11, Ari14},
CAST was configured once again for vacuum running in 2013 by removing the thin
X-ray windows (which had a cutoff at 1$\,$keV). This produced an uninterrupted vacuum
line, running from the vacuum port of the magnet cryostat at one end
of the magnet, through the cold bores of the 10$\,$m prototype LHC magnet,
to the exit port of the cryostat on the opposite side of the magnet.

X-ray detectors on CAST in the period from 2003 to 2012 have operated with
energy thresholds above 1~keV to cover the solar axion energy spectrum.
The 2013 vacuum setup allowed sub-keV photons to exit the magnet
cold bore and reach the X-ray detectors without absorption. Sub-keV
sensitive detectors were then able to explore this energy range.

The experiment described here took place in a short running period 
before the installation of a powerful
combination of the existing X-ray telescope (MPE-Abrixas flight spare) and a
newly developed InGrid detector~\cite{Kri13}, capable of simultaneous sub-keV
and multi-keV operation. A sub-keV detector system
was assembled using mostly commercially available equipment to exploit
this first period of vacuum running. The detector was installed on
the sunrise side of the experiment taking data during the morning
solar tracking of the magnet for $\sim\!90\,$min each day. The magnetic
field length was 9.26$\,$m, the cold bore diameter 43$\,$mm and the
field 9$\,$T.

\section{The detector system}

The X-ray detector system comprised a Silicon Drift Detector (SDD)~\cite{Gat84} and
a preamplifier-readout card$\,$\footnote{Readout Electronics Board (pulsed reset) from PNDetector, Munich, Germany.} inside a vacuum enclosure.
The 1.1$\,$W dissipated from the preamplifier was removed
by a copper heat exchanger block. The SDD signal
was routed to a Digital Pulse Processor (DPP)$\,$\footnote{PX5-Digital Pulse Processor AMPTEK, Bedford, USA.}. The DPP was operated
with a peaking time of \SI{5.6}{\micro\second} in gated mode using the gate provided
by the preamplifier-readout card. The energy threshold of the
device was set to 167$\,$eV.

\begin{figure}[tbh]
\centering{}\includegraphics[scale=0.36]{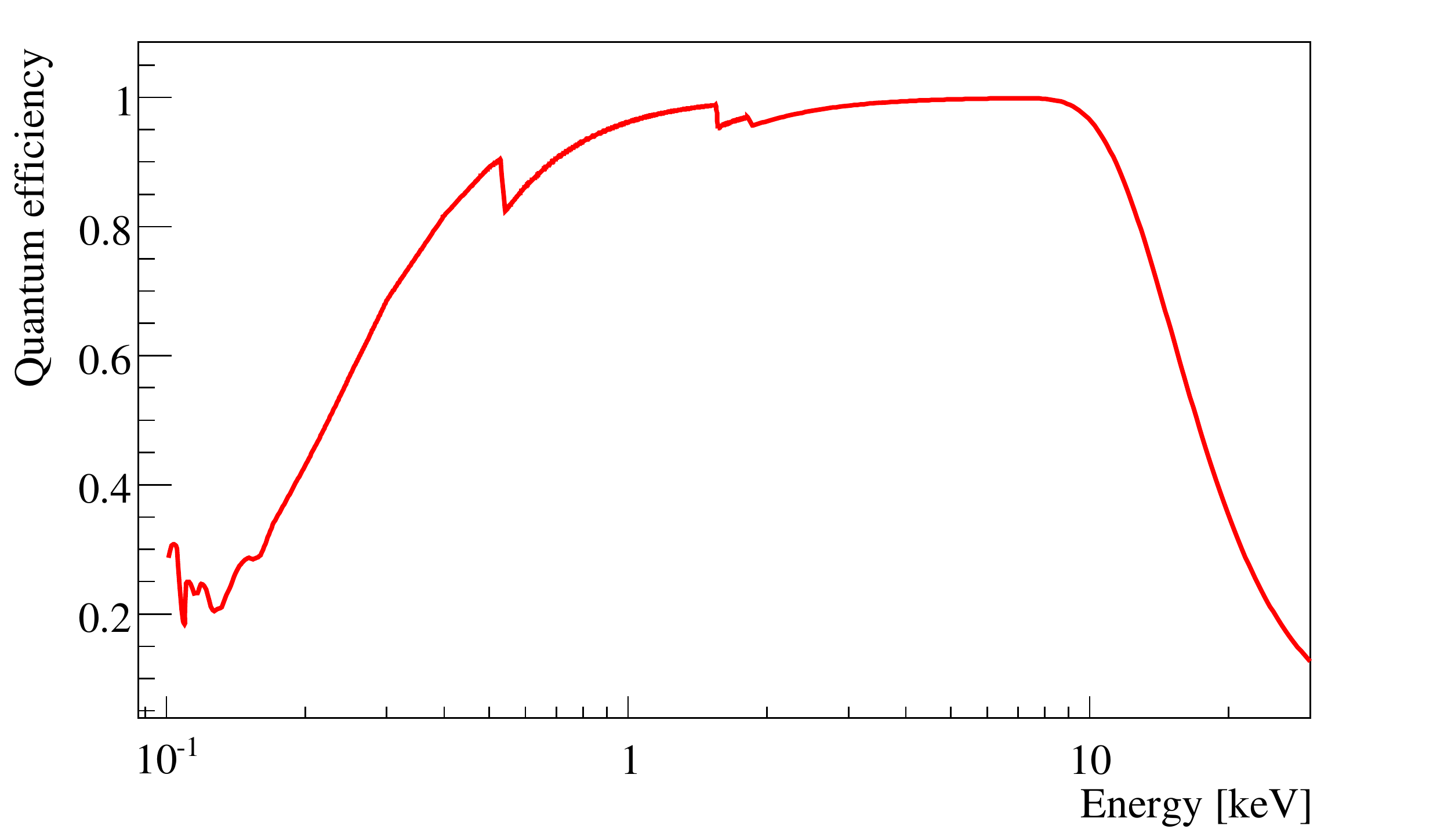}\protect\caption{Quantum efficiency of the SDD.}
\label{QE}
\end{figure}

The detector chosen was a single channel, non-imaging
SDD, without a vacuum window, in this case a commercial research grade device$\,$\footnote{SDD-100-130pnW-OM-ic Premium Line from PNDetector, Munich, Germany.},
with a large surface of 89$\,$mm$^{2}$ effectively covering 6.13\% of the magnet
cold bore (diameter 43$\,$mm). The device was made from \SI{450}{\micro\meter}
thick polysilicon technology with an entrance window optimized for light elements. The energy resolution
of the device is 39$\,$eV FWHM at 277$\,$eV. Due to the sharply rising noise profile at low energies,
 the threshold used in the analysis was set
at 400$\,$eV. The typical quantum response is shown
in Fig.~\ref{QE}; the quantum efficiency ($\varepsilon_{q}$) exceeds 80\% for photon energies above 400$\,$eV.
The background level for the device was $\sim\!\!10^{-3}\,$cts/keV/cm$^{2}$/s in the range $400-1500$~eV 
and was independent of the detector temperature over the range $-25^{\circ}$ to $-45^{\circ}$C.
The SDD was operated at $-30^{\circ}$C using an integrated double
peltier cooling element; the 1.0$\,$W from the SDD was also removed by the copper
heat exchanger. The detector temperature remained constant when tilting the magnet during solar tracking.

\begin{figure}[tbh]
\begin{centering}
\includegraphics[scale=0.37]{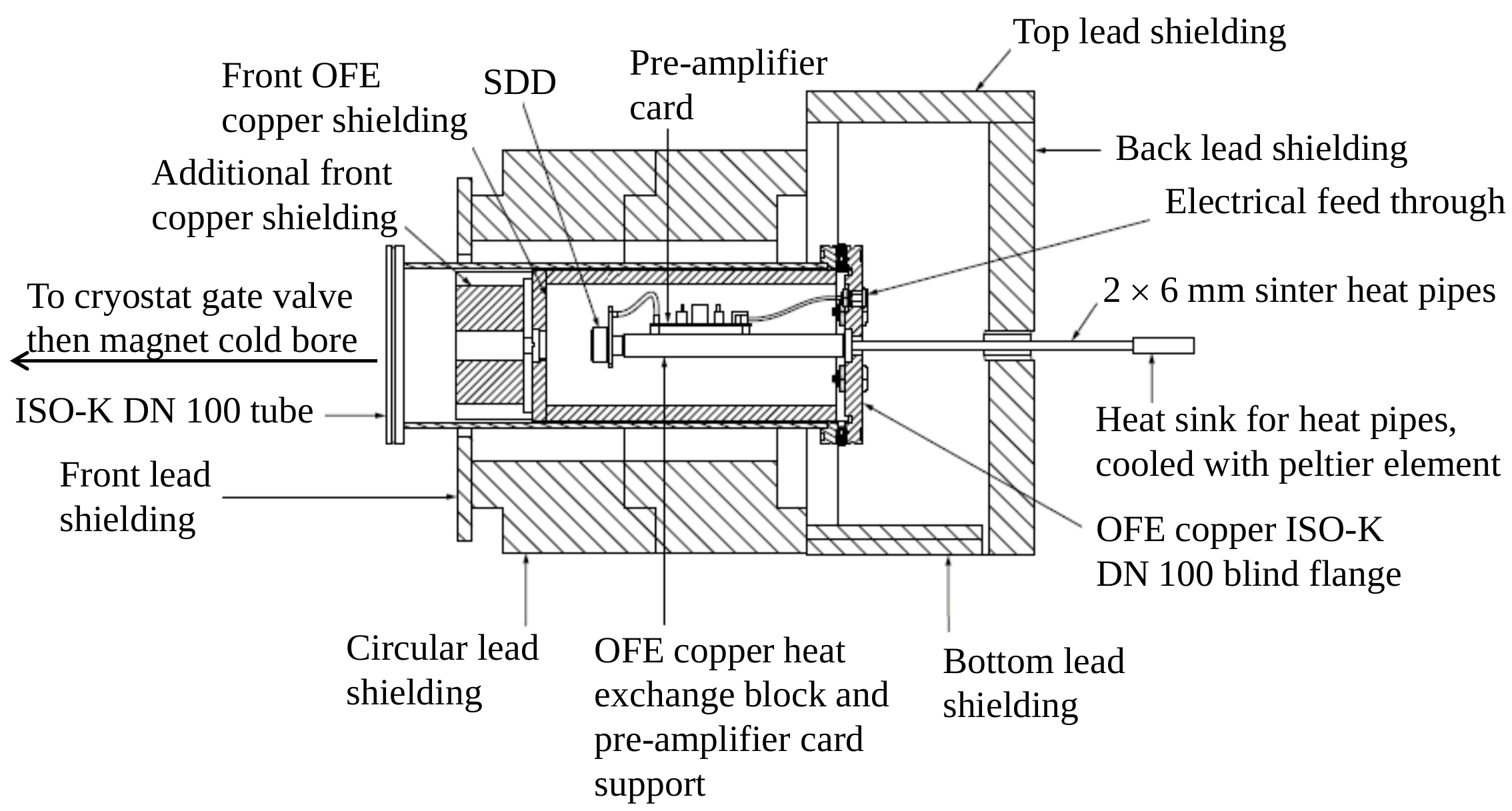}\protect\caption{SDD detector vacuum and shielding system.}
\par\label{vac_shield}\end{centering}
\end{figure}

The detector system inside the vacuum vessel was connected directly to the
cold bore vacuum port gate valve (on the left in Fig.~\ref{vac_shield}). The vacuum vessel was made from an iso-K
DN 100 stainless steel tube connected to a custom-built copper end
flange.

Shielding inside the vacuum vessel was provided by the OFE copper
back flange plus an OFE copper inner cylinder and upstream collimator;
the vacuum vessel was surrounded by 6$\,$cm thick lead rings, and lead
plates with thickness between 1 and 3$\,$cm. The turbo-pumped dry vacuum
system operated at a pressure of $1.2\times10^{-6}\,$mbar with the cryostat
gate valve opened for data taking.

\section{Laboratory tests}

\begin{figure}[tbh]
\begin{centering}
\includegraphics[scale=0.32]{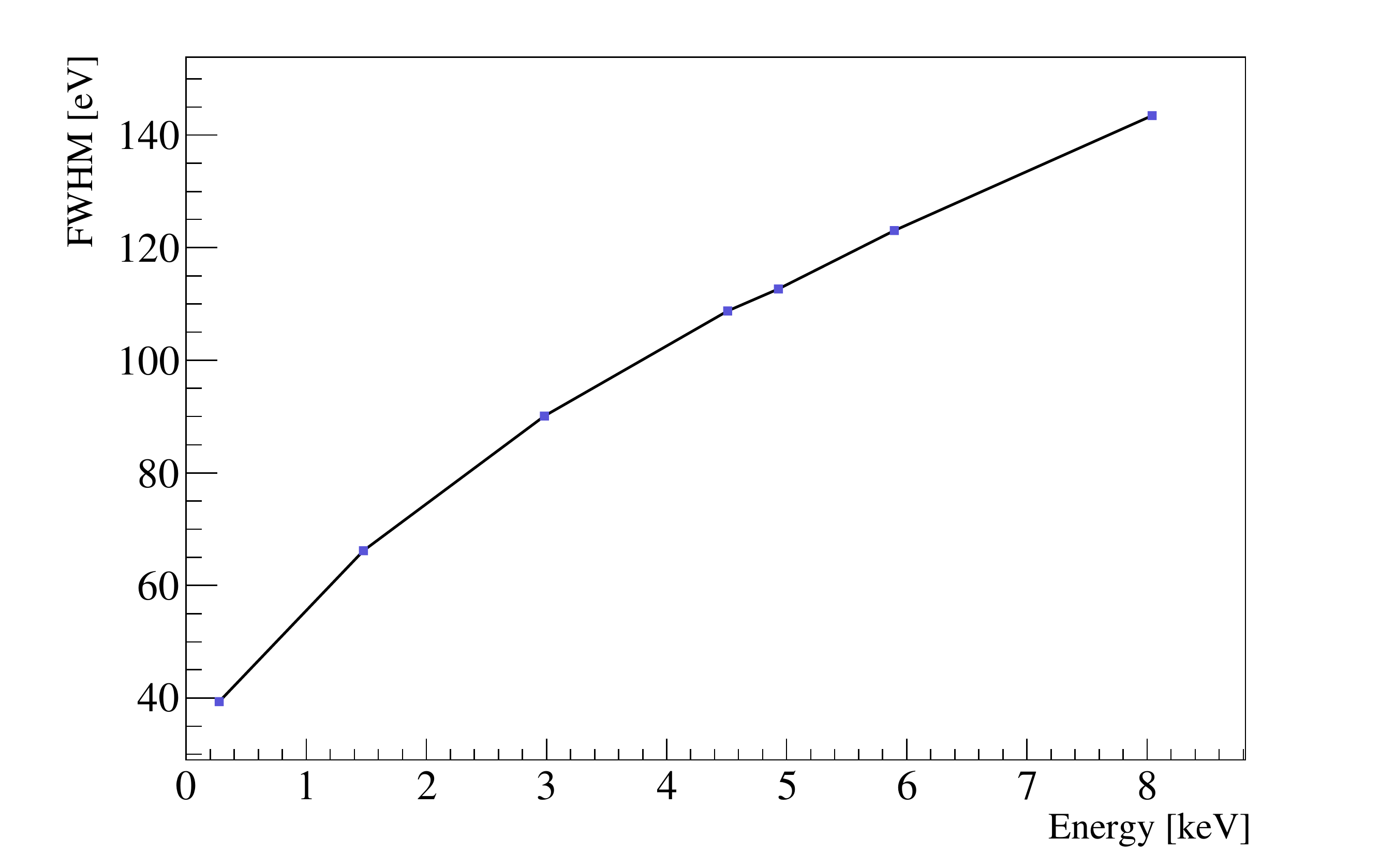}\protect\caption{Measured FWHM of the SDD versus energy.}
\par\label{calib}\end{centering}
\end{figure}

Prior to installation on CAST, the SDD was tested in a laboratory
at CERN on a variable energy X-ray vacuum beam line~\cite{Vaf12}.
The system provided calibration energies between 0.28 and 10$\,$keV.
Using the characteristic emission lines with
the best statistics, the energy resolution (FWHM) of the detector
was defined for various energies. In Fig.~\ref{calib} the measured FWHM versus the energy is shown.

During the first run of the detector on the X-ray beam line, a noticeable
drop in the count rate with time was observed in the energy range $300-800\,$eV
with the detector operated at $-46^{\circ}$C and in a relatively
poor vacuum of $4\times10^{-5}\,$mbar.
To quantify this phenomenon further, measurements were taken
at the nominal detector operating temperature of $-30^{\circ}$C, using the
bremsstrahlung spectrum of the Ag target in the X-ray generator.
After a few hours at room temperature, the detector was cooled to
$-30^{\circ}$C at a vacuum pressure of $2.5\times10^{-6}\,$mbar. 
Then several 5$\,$min measurements of the same
spectrum were taken over the course of the next 25$\,$h to determine the loss
of efficiency with time. Returning the detector to room temperature for 1$\,$h 
was enough to fully recover its efficiency.

The loss of efficiency was attributed to substances outgassing from the materials 
inside the vacuum system  and then being cryo-pumped on to the cold entrance window surface. 
Both the experimental and the laboratory test setups used standard, surface-cleaned 
stainless steel High Vacuum (HV) components and dry Viton joints. The SDD preamplifier card, 
whilst intentioned for vacuum use, was not constructed from HV materials 
and components. The wiring and connectors between SDD, preamplifier card and the 
electrical vacuum feed-throughs were not HV standard.
As these components were common to both systems (laboratory and experimental), 
the resulting loss of efficiency measured in the laboratory is believed to have also been 
present in the experiment.

\begin{figure}[tbh]
\includegraphics[scale=0.38]{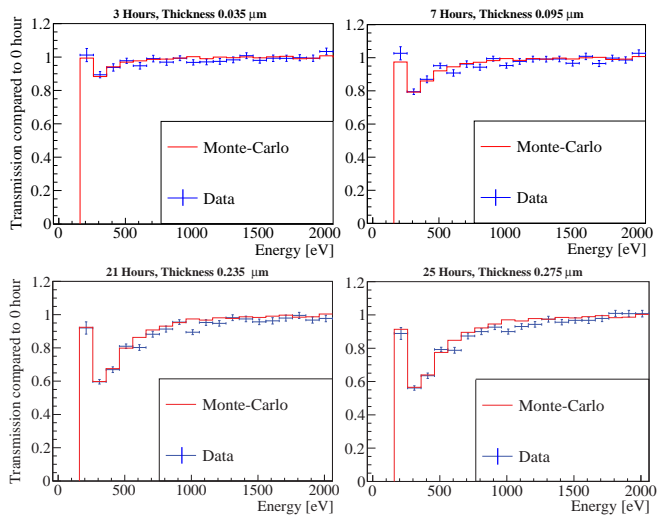}\protect\caption{Comparison of the 
spectrum that was taken immediately after the cool-down of the SDD, 
with the ones taken $3,\,7,\,21$ and $25\,$h later. The form of the histograms 
clearly indicates a progressive deposition of an absorption layer on the detector surface. The simulated
data (continuous line) correspond to the deposition of a C$_3$H$_6$ film
on the surface of the detector.}
\label{transmiss1}
\end{figure}

\begin{figure}[tbh]
\begin{centering}
\includegraphics[scale=0.36]{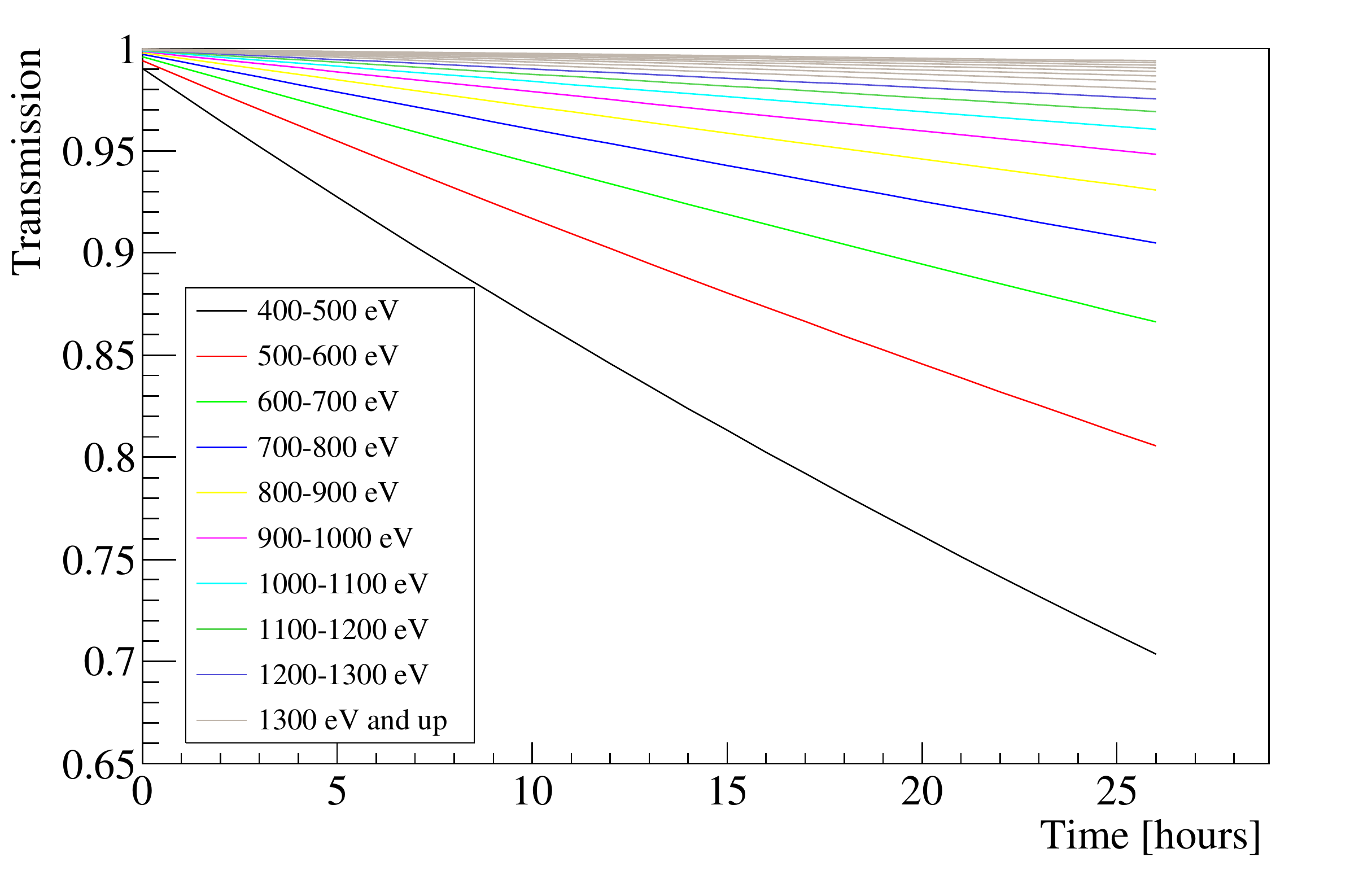}\protect\caption{The drop in transmission due to the increasing deposition of the hydrocarbon
film on the surface of the detector, for each energy bin, at the energy
range of interest ($400-1500\,$eV). }
\par\label{transmiss2}\end{centering}
\end{figure}

Monte-Carlo simulations verified that the loss in efficiency of the
SDD could be explained by a film deposition with a thickness that
increases uniformly with time. Moreover the absorption spectrum of
a simple hydrocarbon film (C$_3$H$_6$) proved to be sufficient when
folded with the resolution of the detector to reproduce the measured
absorption. In Fig.~\ref{transmiss1} the comparison of the spectra taken after 3, 7, 21
and 25$\,$h of operation of the detector, with the initial one
(directly after cool down) are displayed, together with the simulated
values. A chi-squared fit was performed between the simulated and
real data to determine the evolution of the thickness of the hydrocarbon
film. The result is shown in Fig.~\ref{transmiss2}, where the evolution of the transmission
of the hydrocarbon film is shown for each bin of 100$\,$eV. These data
were used for the parameterization of the transmission of the hydrocarbon
film, versus time and energy, in order to calculate the overall efficiency
of the detector to photons in the range $400-1500\,$eV. Our correction 
for this effect to the tracking data is less than 3\%. 
Our SDD proved to be significantly
more efficient than an SDD fitted with a vacuum window.

\section{Data taking}

The tests in the laboratory indicated that the detector required 1$\,$h 
at ambient temperature in order to fully recover its lost
efficiency. To ensure maximum efficiency of the detector during the
sunrise solar tracking, the detector was set to ambient temperature 2$\,$h before
data taking and set to $-30^{\circ}$C only 30$\,$min before the sunrise
solar tracking started. At the end of the sunrise solar tracking the detector
was again set to ambient temperature and then set back to $-30^{\circ}$C about
30 min before the evening solar tracking. The detector then remained
at nominal operating temperature until the next day, 2$\,$h before sunrise solar tracking,
when the cycle was repeated. The data taking took place over 9 sunrise solar trackings amounting
to 15.2$\,$h of exposure. The background data consisted of 13.8$\,$h of sunset solar
tracking and 94.2$\,$h of overnight background runs with the magnet stationary (108$\,$h in total).

\begin{figure}[tbh]
\centering{}\includegraphics[scale=0.36]{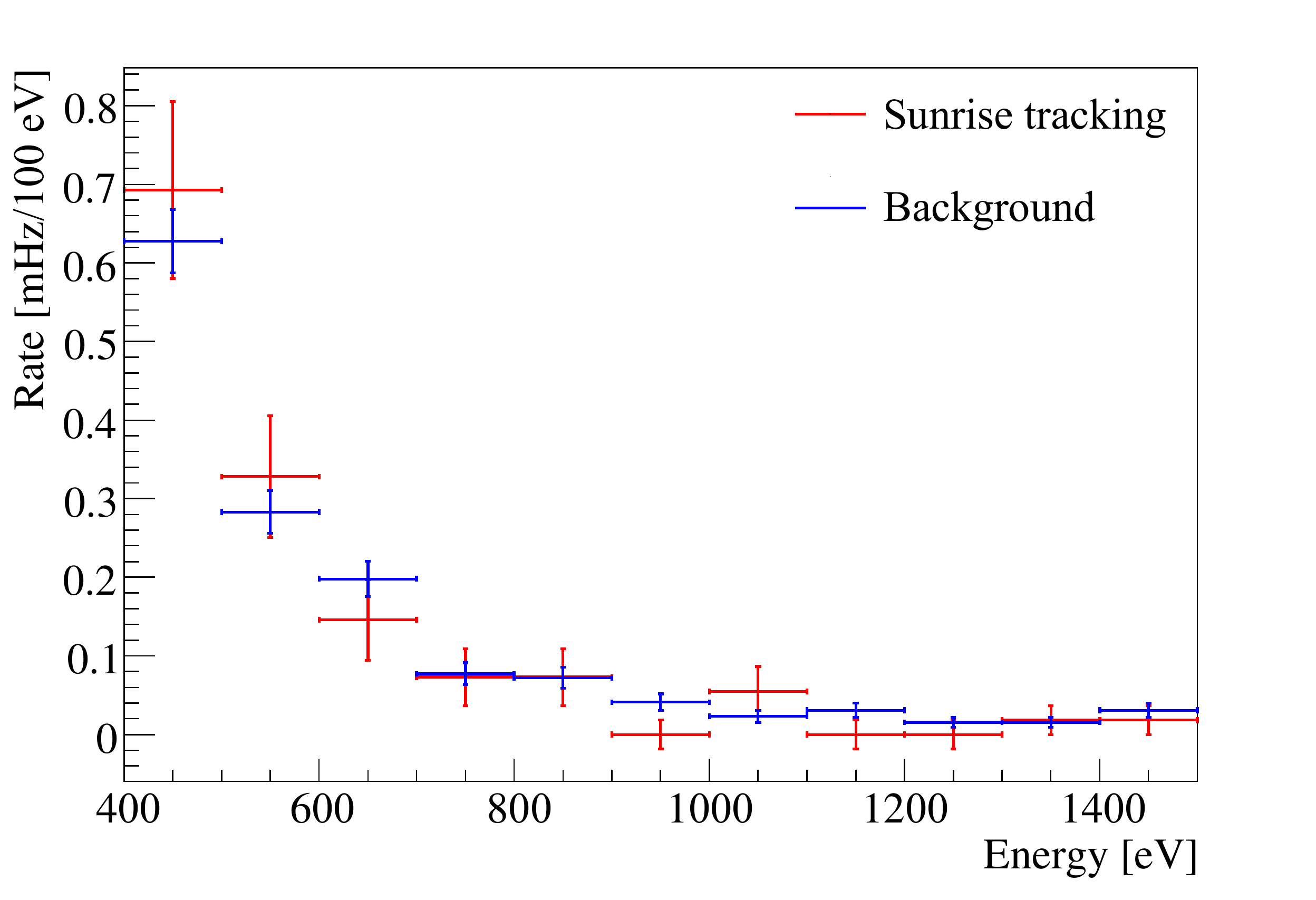}\protect\caption{Combined spectra of the rate during sunrise tracking and during background measurements.}
\label{spectra}
\end{figure}

The operational energy threshold for the SDD was 167$\,$eV which produced
an acquisition rate of $\sim$5$\,$mHz over the range up to 10$\,$keV.
This rate was quasi-constant and independent of the magnet motion.
Over the whole data taking period of 9 days the SDD rate
between 400 and 1500$\,$eV was 1.40$\pm$0.16$\,$mHz (15.2$\,$h
sunrise tracking) corresponding to 1.43$\,\times\,$10$^{-3}\,$cts/\,keV/\,cm$^{2}/\,$s.\newline
The rate obtained during background runs was 1.42$\pm$0.06$\,$mHz
(108$\,$h). The spectra of the sunrise tracking and background
rates are shown in Fig.~\ref{spectra}.

\begin{figure}[tbh]
\centering{}\includegraphics[scale=0.33]{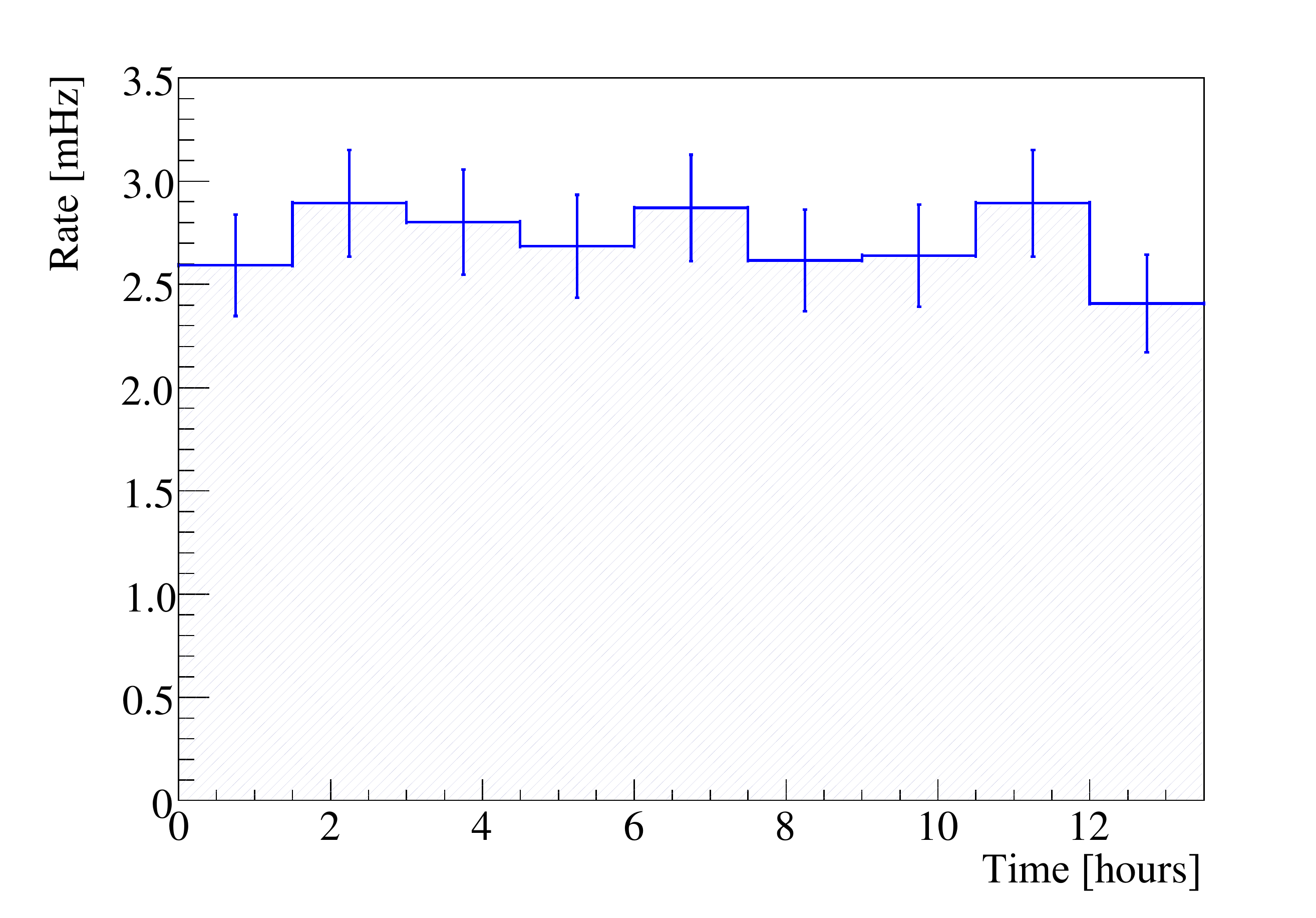}\protect\caption{Evolution of low energy background ($300-600\,$eV) with time.}
\label{timedep}
\end{figure}

The effect of the internal copper and external lead shielding in the background can
be gauged by the comparison between data taken in the X-ray laboratory
(unshielded) and from CAST (both internal copper and external lead
shielding). The background rate in the range $1.5-10\,$keV of the SDD in the X-ray
beam line was 3.86$\pm$0.34$\,$mHz (compared to 1.42$\pm$0.06$\,$mHz on CAST) 
and for $400-1500\,$eV  was 2.31$\pm$0.26$\,$mHz (compared to 1.42$\pm$0.06$\,$mHz
on CAST), indicating the presence of electronic noise at low energies.
Analysis of the energy spectra for all background runs taken over
typically 13.5$\,$h each day showed no statistically significant decrease with
time in the spectra at low energies ($300-600\,$eV), as shown in Fig.~\ref{timedep}.

\section{Theoretical chameleon spectrum}

\def\be{\begin{equation}} \def\ee{\end{equation}} 
Chameleons can be produced by mixing with the photon flux emanating from the Sun's core~\cite{Bra10,Bra12}. The conversion probability of photons into chameleons in a magnetised region with a constant magnetic field $B$ over a distance $l$ is given by~\cite{Bra12}
\begin{equation}
p_{\gamma\to\phi}(l)= \frac{\beta_\gamma^2 B^2l_\omega^2}{4 m_{\rm Pl}^2 }\sin^2 \frac{l}{l_\omega}\ ,
\label{prob}
\end{equation}
where the Planck mass is $m_{\rm Pl}\!\!\sim\!\!2\times 10^{18}$~GeV, the coherence length is given by $l_\omega= \frac{4\omega}{m^2_{\rm eff}}$ and the effective mass of the chame\-leon is
\begin{equation}
m^2_{\rm eff}=\beta_{\rm m}^{(n+2)/(n+1)} \omega_\rho^2 -\omega_{pl}^2\ ,
\end{equation}
where we have defined 
\begin{equation}
\omega_\rho^2= \frac{(n+1) \rho}{m_{\rm Pl}} (\frac{\rho}{n\,m_{\rm Pl} \Lambda^{n+4}})^{1/(n+1)}
\end{equation}
and the plasma frequency is $ \omega_{pl}^2 = \frac{4\pi \alpha\rho}{m_em_p}$.
We have introduced the fine structure constant $\alpha\!\sim\!1/137$ and the proton and electron masses $m_p$ and $m_e$. The mass of the chameleon depends on the density $\rho$ and the coupling to matter $\beta_{\rm m}$. The index $n>0$ defines the chameleon model and comes from the scalar potential $\frac {\Lambda^{n+4}}{\phi^n}$ where $\Lambda\!\sim\! 10^{-3}$~eV is the dark energy scale.
We have assumed that the mixing angle $\theta= \frac{\omega B \beta_\gamma}{m_{\rm Pl} m^2_{\rm eff}} \lesssim 1$.

Photons in the solar plasma perform a random walk. When they have moved by a radial distance $d(l)$ in one second, they have undergone $N(l)$ collisions with the plasma where $l$ is the distance between two collisions
\be
N(l)= \frac{c}{d(l)},\ \ d(l)= l\sqrt{N(l)}\ .
\ee
The distance $l$ is distributed according to a Poisson law with average $\lambda$ given by the mean free path. In a solar region of width $\Delta R$ where the mean free path and the magnetic field are (nearly) constant, the conversion probability into chameleons is given by
\be
d{\cal P} (l)=\frac{\Delta R}{d(l)} N(l) p_{\gamma\to\phi}(l) e^{-l/\lambda} \frac{dl}{\lambda}\ .
\ee
Summing over the total number of cells defined by $R_\odot/\Delta R$ we get the conversion rate per unit length
\be
\!\!\!\!\!\!\!\frac{d{\cal P}}{dx}= \\ \sqrt{\frac{c}{l_\omega(r)}} \frac{\beta_\gamma^2 B^2(r)l_\omega^2(r) R_\odot}{4m_{\rm Pl}^2 \lambda (r)}\int_0^\infty \frac{\sin^2 y}{y^{3/2}} e^{-l_\omega (r)y/\lambda(r)}dy \ ,
\ee
which depends on the radius $r$ from the centre of the Sun. The conversion probability is obtained by integrating the conversion rate over $x=r/R_\odot$.

In the tachocline, and for the range of energies of interest it turns out that $l_\omega (r)\ll \lambda (r)$, implying that the conversion rate simplifies greatly
\be
\frac{d{\cal P}}{dx}= C \sqrt{\frac{c}{l_\omega(r)}} \frac{\beta_\gamma^2 B^2(r)l_\omega^2(r) R_\odot}{4m_{\rm Pl}^2 \lambda (r)} \ ,
\ee
where $C=\int_0^\infty \frac{\sin^2 y}{y^{3/2}}dy $. Notice that the spectrum depends on $\omega^{3/2}$ and not $\omega^2$ due to the random walk of the photons in the solar plasma, and the $\sqrt{N(l)}$ excursions of the photons covering the distance $d(l)$ in one second.
In practice and in the absence of a resonance where $m^2_{\rm eff}$ vanishes somewhere in the tachocline for a large value of $\beta_{\rm m}$, the effective mass of the chameleon is essentially independent of the coupling to matter $\beta_{\rm m}$.
This implies that the conversion probability depends only on the coupling to photons, $\beta_\gamma$.

The chameleon flux leaving the Sun is simply given by
\be
\Phi_{\rm cham}(\omega)= \int_{0}^1 n_\gamma p_\gamma \frac{d{\cal P}}{dx} dx \ ,
\ee
where the integrand vanishes outside the tachocline. It depends on the photon flux $n_\gamma$ and the photon spectrum $p_\gamma$. The spectral dependence of this flux is in $\omega^{3/2} p_\gamma(\omega)\sim \frac{\omega^{7/2}}{e^{\omega/T}-1}$, where $T$ is the photon temperature in the tachocline, with a maximum around $\omega_{\rm max}\!\sim\!600\,$eV.
The total luminosity of the Sun in chameleons is given by
\be
L_{\rm cham}=\int_0^\infty  \omega \Phi_{\rm cham}( \omega) d\omega \ ,
\ee
which depends on $\beta_\gamma^2$. We calibrate $\beta_\gamma$ in such a way that the chameleon luminosity does not exceed 10\% of the solar luminosity.
 For $n=1$ and a tachocline of width $0.01\,R_\odot$ located at a radius $0.7\,R_\odot$ and a magnetic field of 10$\,$T, the chameleons saturate the solar luminosity bound for
 $\beta_{\gamma}^{\rm sun}=10^{10.81}$. As the number of regenerated photons in the CAST detector is proportional to $\beta_\gamma^4$, this gives an upper limit to the number of photons that one may expect to detect.
 In the following, we shall see how the likelihood analysis takes into account the solar bound on the chameleon luminosity.

\begin{figure}[tbh]
\begin{centering}
\includegraphics[scale=0.35]{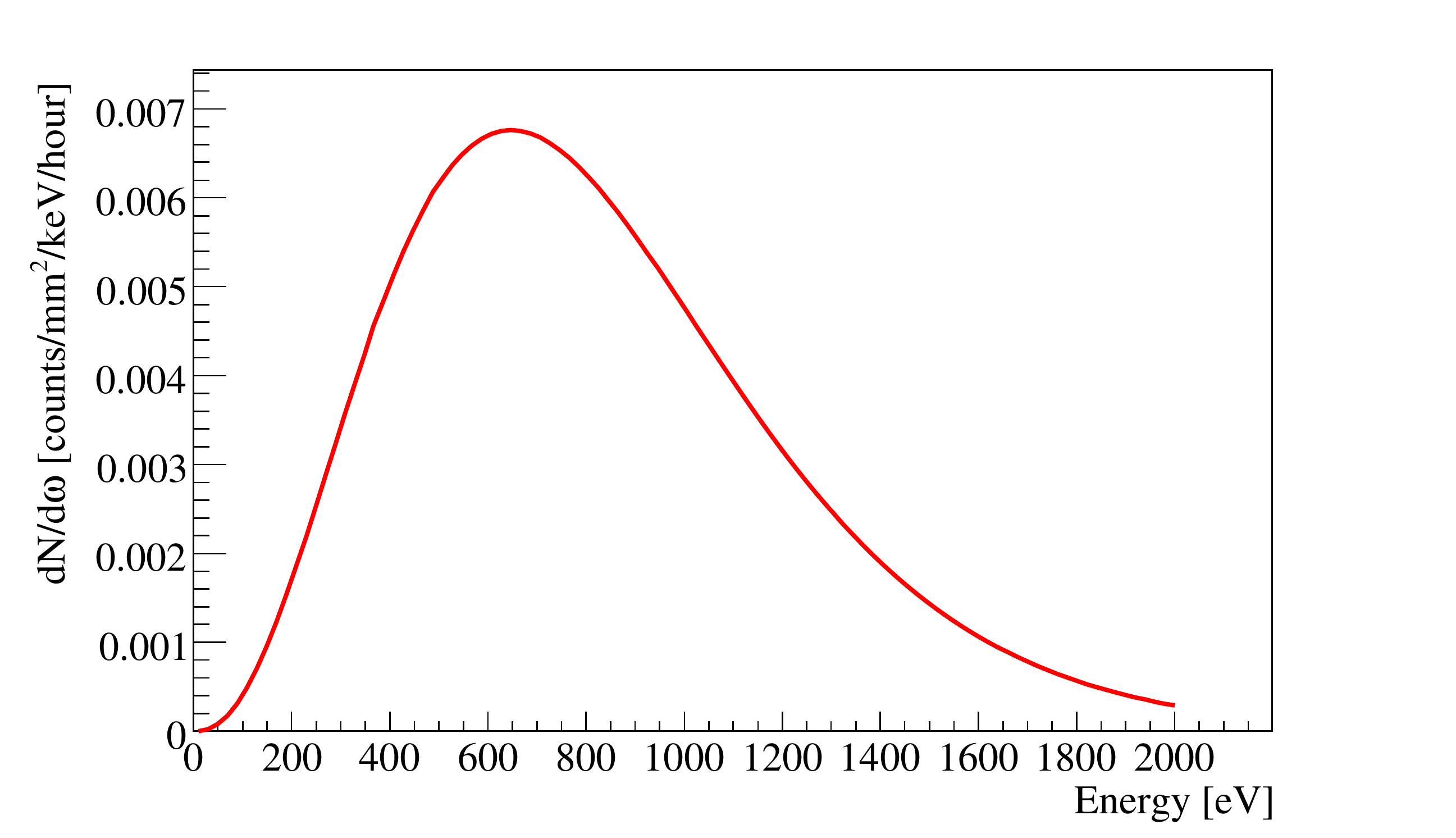}\protect\caption{{Expected number of photons arriving at the SDD, for $\beta_{\gamma}^{\rm sun}=10^{10.81}$, assuming all chameleons pass through the full magnetic length of the CAST magnet and assuming no absorbing material upstream of the cold bore.} }
\label{theor_ph}
\par\end{centering}
\end{figure}

\section{Analysis and results}

The cold bore diameter of 43$\,$mm at the upstream end of the magnetic
region of $L_{0}$=9.26$\,$m results in an aperture of 3.5$\,$mrad as seen
by the SDD. Chameleons emitted from larger angles up to the tachocline
(6.5 mrad for a sphere diameter of $0.7\,R_\odot$) traverse less than the full magnetic length of the magnet. A~simulation 
of the CAST geometry was carried out and the results
were that 15.7$\%$
of emitted chameleons passed through the full
field length, the remainder pass through varying lengths ($L$) which when
integrated and scaled by the $(L/L_{0}$)$^{2}$ factor are equivalent to
a further 23.2$\%$ passing through the full length.
In total a scale factor ($F$) of 38.9$\%$ has to be applied to the expected number of photons to
account for the fact that not all chameleons that reach the detector
pass through the full magnetic length. 

All chameleons from the tachocline incident on the SDD must pass via the 
lead shielding on the sunset side of the magnet before entering the magnet cold bores. 
The 400 eV energy threshold in the analysis is well above the maximum chameleon 
effective mass in lead ($m_{\rm eff}$=135$\,$eV for $n$=1, $\beta_{\rm m}=10^{6}$), 
hence no absorption effects occur within our region of interest.

\begin{figure}[tbh]
\centering{}\includegraphics[scale=0.34]{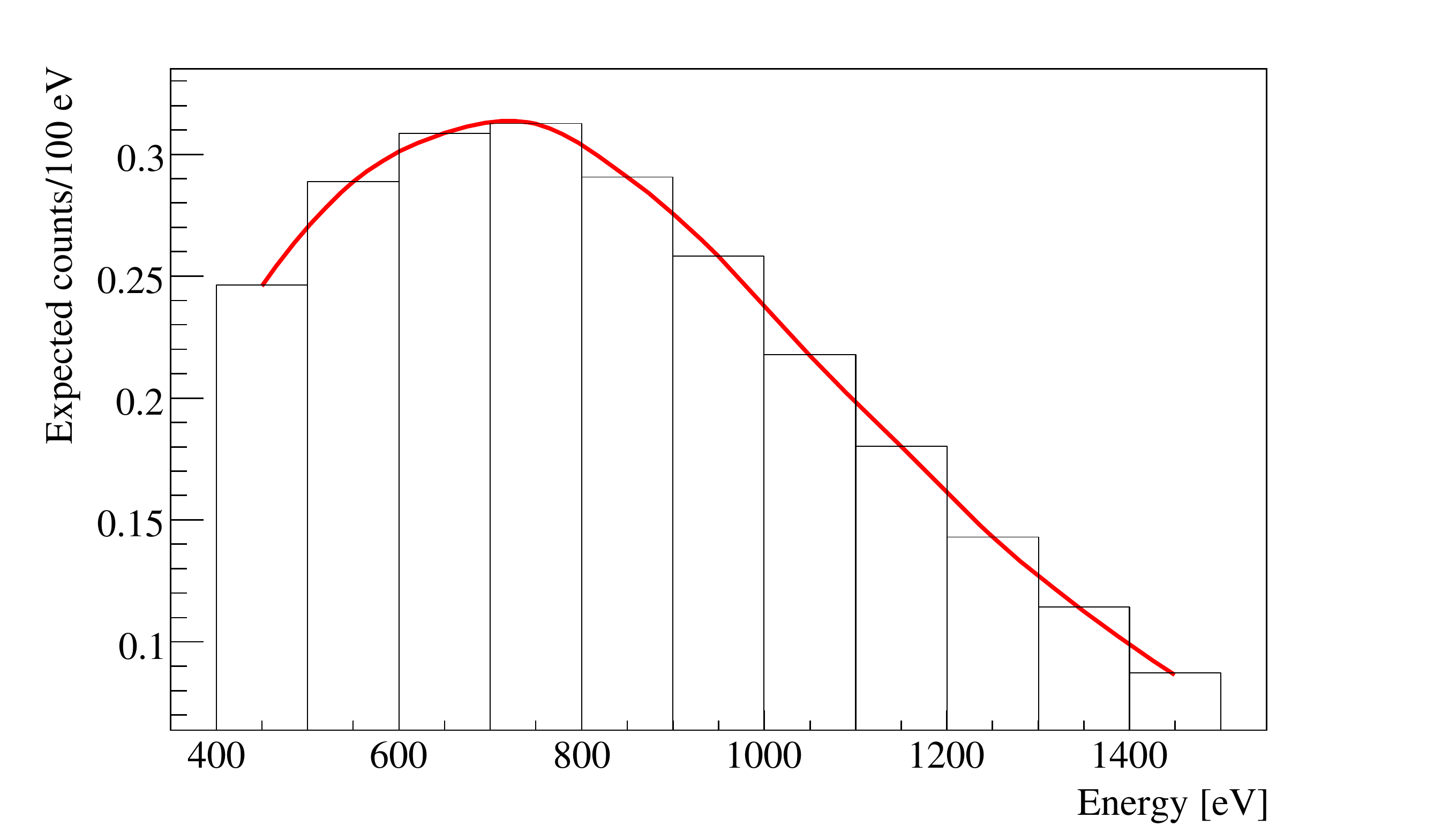}\protect\caption{Expected number of photons ($\beta_{\gamma}^{\rm sun}=10^{10.81}$) to be detected by the SDD taking into account the total tracking time. }
\label{exp_photons}
\end{figure}

The expected number of photons from chameleon conversion inside the
CAST magnet, that will reach the SDD is calculated from the theoretical
photon spectrum (Fig.~\ref{theor_ph}) arriving at our detector taking into account the total
tracking time, the quantum efficiency of the detector, the magnetic
length that the chameleons travel inside CAST, the absorption phenomena
on the surface of the SDD and the area of the detector:

\be
N_{i}^{\rm ch}=f\left(E_{i},\beta_{\gamma}^{4}\right)\times A_{\rm SDD}\times t\times F\times\varepsilon_{q}\times\varepsilon_{\rm abs}\times dE \ ,
\label{Ni}
\ee
where the index $i$ runs over the energy bins, $f\left(E_{i}\right)$ is the expected number of photons given in cts/100 eV/mm$^{2}$/s
in front of our detector, calculated with $\beta_{\gamma}^{\rm sun}=10^{10.81}$
and having travelled the full length of the CAST magnet, $A_{\rm SDD}$ the area of the detector,
$t$ the total tracking time in seconds, $\varepsilon_{q}$ the quantum efficiency of the detector,
$\varepsilon_{\rm abs}$ the transmission of the thin absorbing layer, that has accumulated
after 2$\,$h on its surface, and $dE$ the energy bin size. The resulting spectrum is shown in Fig.~\ref{exp_photons}.

The analysis of the data has been performed by using the likelihood method.
 For data that follow a Poisson distribution the likelihood function can be expressed as

 \be
\log(L)=\Sigma_{i}\left(-\lambda_{i}+t_{i} \, \log\left(\lambda_{i}\right)-\log\left(t_{i}!\right)\right) \ ,
\ee
 where $t_i$ is the number of tracking counts  in the energy bin $i$ and
 $\lambda_{i}$ is the expected number of counts:
\be
\lambda_{i}=b_{i}+N_{i} \, C \ ,
\ee
with $b_{i}$ the expected background in energy bin $i$ and $N_{i} \, C$
the expected number of photons from chameleon conversion, which is proportional to the quantity $\beta_{\gamma}^{4}$ (eq.~\ref{Ni} ). For
simplicity we choose the free parameter as $C=\left(\beta_{\gamma}/\beta_{\gamma}^{\rm sun}\right)^{4}$
to evaluate the data.

\begin{figure}[tbh]
\centering{}\includegraphics[scale=0.36]{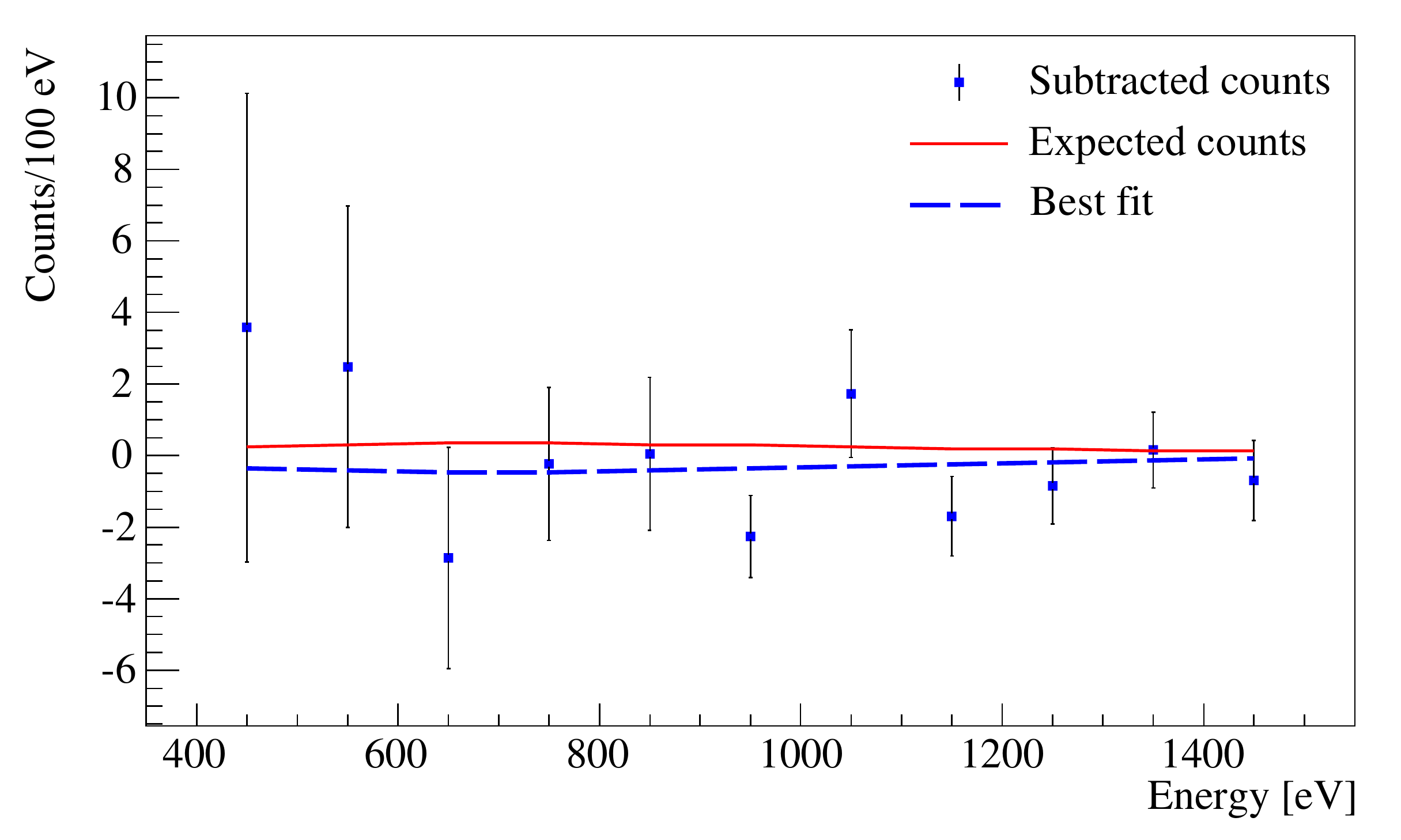}\protect\caption{Subtracted counts, expected number of counts during the solar tracking (red) and best fit to the data from the maximisation of the Likelihood (blue).}
\label{substract}
\end{figure}

The maximum of $\log(L)$ will be achieved by tuning
the parameter $C$. The obtained value $C_{\rm Best\, fit}$ is compatible with the null hypothesis within 1 sigma. In
Fig.~\ref{substract} the subtracted counts, tracking minus background (normalised to tracking time),
together with the expected photon signal from chameleon conversion,
and the best fit to our data are shown.

The upper limit on $\beta_\gamma$ is then obtained by integrating the Bayesian probability
with respect to $C$ from 0 up to $95\%$, considering only the non-negative
part of the distribution. The resulting bound on  $\beta_{\gamma}$ is
\be
\beta_{\gamma}\leq 9.26\times10^{10}\ {\rm at}\ 95\% \ {\rm CL} \ .
\ee

\begin{figure}[tbh]
\centering{}\includegraphics[scale=0.36]{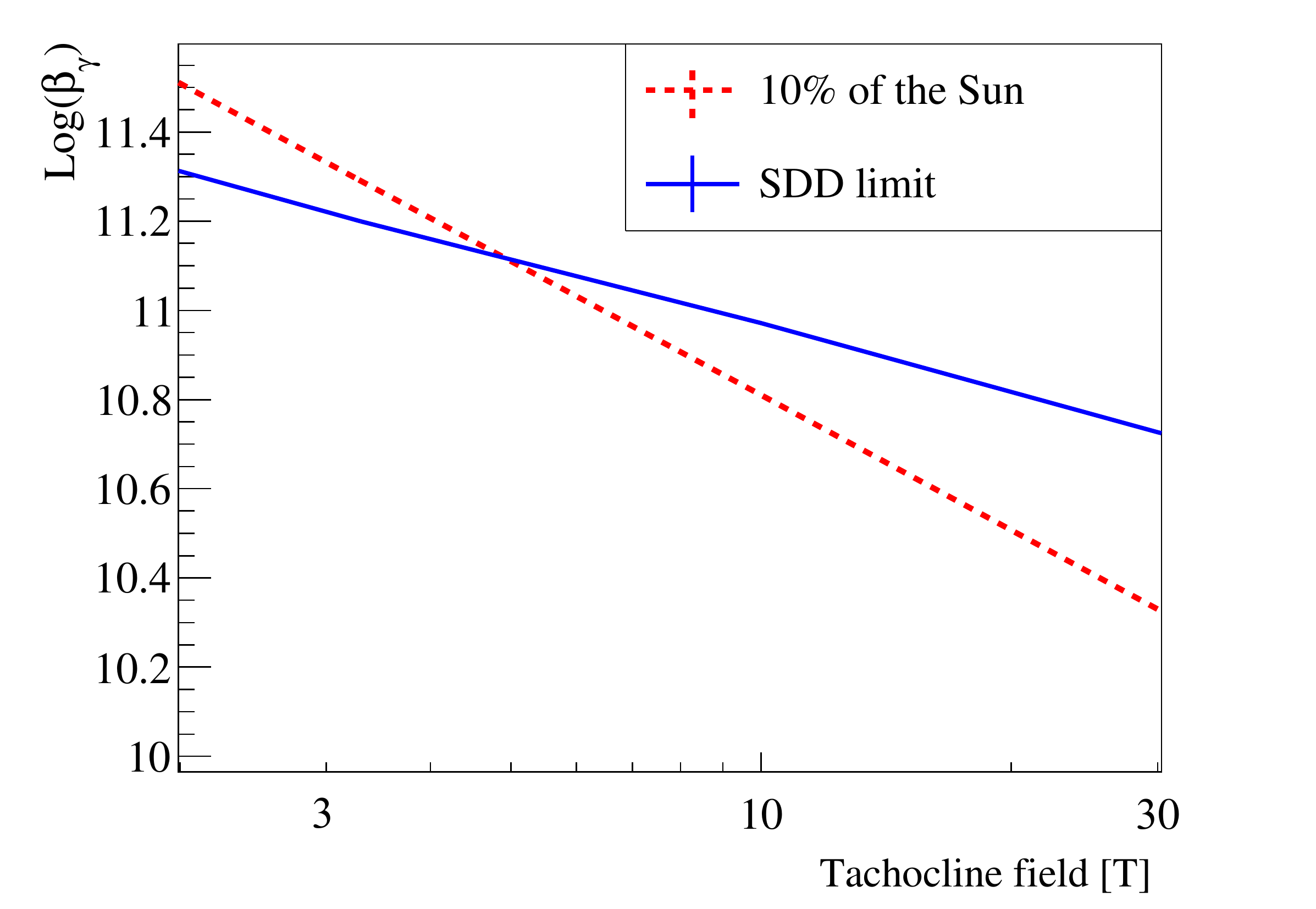}\protect\caption{The CAST limit on $\beta_\gamma$ for different values of magnetic field in the tachocline.}
\label{Allbeta}
\end{figure}

Our result can be modulated depending on the type of solar model considered.
Indeed, we have focused on the $B=10\,$T case in the tachocline. The uncertainty on the tachocline 
field is believed to be in the range 4 to 25-30$\,$T~\cite{Web13,Cal95,Ant03}. Hence the CAST limit on the photon coupling can be shifted by
a factor of about plus or minus $2.5^{1/2}$ as can be seen in  Fig.~\ref{Allbeta}.
The limit obtained with the SDD is actually lower than the solar luminosity bound for values
 of tachocline magnetic fields below 4.9$\,$T.

Additionally, we have shifted the position of the tachocline down  to $0.66\,R_\odot$ and increased its width from $0.01\,R_\odot$ to $0.04\,R_\odot$. We have also
considered a linearly decreasing magnetic field (10$\,$T at $0.7\,R_\odot$ down to 0$\,$T at $0.8\,R_\odot$). The changes to the bound on $\beta_\gamma$ can be found in Table~\ref{tab1}. On the whole and irrespectively of the astrophysics of the tachocline, we have found that the coupling of photons to chameleons satisfies $\beta_{\gamma}\le 10^{11}$.

\begin{table}[tbh]
\begin{centering}
\begin{tabular}{|c|c|c|c|}
\hline
Tachocline [$\odot$] & Width [$\odot$] & $\beta_{\gamma}$ at $95\%$ CL & $\beta_{\gamma}^{\rm sun}$ \tabularnewline
\hline
\hline
0.66 & 0.04 & 5.69$\times$10$^{10}$ & 2.95$\times$10$^{10}$ \tabularnewline
\hline
0.66 & 0.01 & 8.9$\times$10$^{10}$ & 5.89$\times$10$^{10}$ \tabularnewline
\hline
0.7 & 0.1 linear & 7.29$\times$10$^{10}$ & 3.47$\times$10$^{10}$ \tabularnewline
\hline
\end{tabular}
\par\end{centering}
\begin{centering}
\protect\caption{Upper limit on $\beta_{\gamma}$ derived from our measurements for different solar models, all for 10\% solar luminosity bound.}
\par\label{tab1}\end{centering}
\end{table}

\section{Discussion}

\begin{figure}[tbh]
\centering{}\includegraphics[scale=0.37]{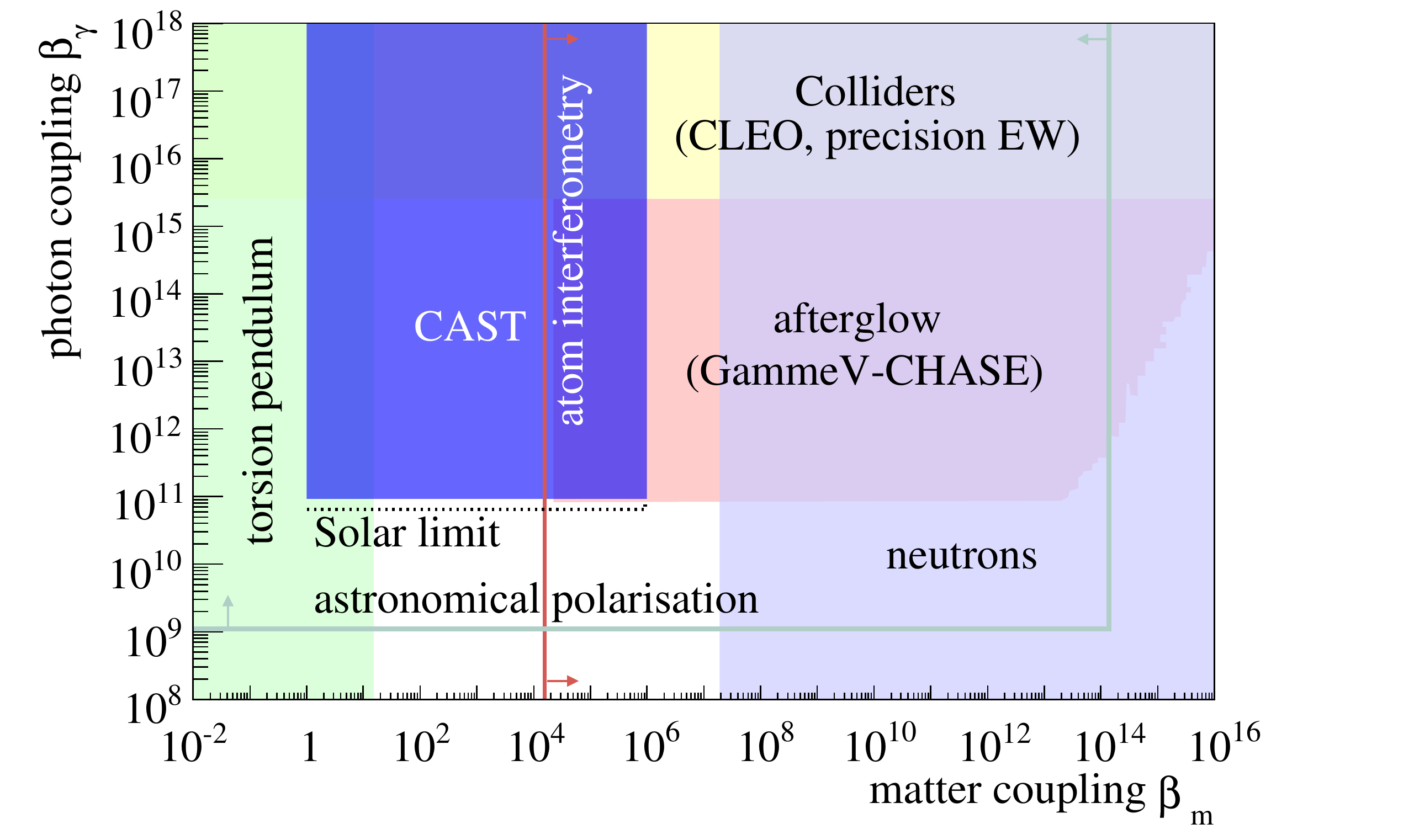}\protect\caption{The exclusion region for chameleons in the $\beta_{\gamma}-\beta_{\rm m}$ plane, achieved by CAST in 2013 (purple). 
We show the bounds set by torsion pendulum tests (in green)~\cite{Upa12}, neutron interferometry measurements (lilac)~\cite{Lem15},
CHASE (pale orange)~\cite{Ste10} and collider experiments (yellow)~\cite{Bra09}. The forecasts of the atom-interferometry technique~\cite{Ham15} and the astronomical polarisation~\cite{Bur08} are represented with lines. }
\label{excl}
\end{figure}

The parameter space of chameleons is determined by the coupling constants
to matter and radiation, and a discrete index $n$ which specifies
the type of dark energy model under consideration. Our result for the upper limit on $\beta_{\gamma}$ is presented in Fig.~\ref{excl},
together with other experimental bounds. A number of experiments are totally
 insensitive to the coupling to photons resulting in vertical lines in the figure. The torsion pendulum tests of the presence of new scalar forces lead to a lower bound
 on the coupling to matter (in green)~\cite{Upa12}. Neutron interferometry tests lead to an upper bound (lilac)~\cite{Lem15}.
 Presently, the atom-interferometry technique is promising the largest reduction in the upper bound \cite{Ham15}
  on the coupling to matter.  Precision tests of the standard model are only
 sensitive to the coupling to gauge fields, i.e. here to photons, and provide
 a large upper bound. From astrophysics, an analysis of the polarisation of the light coming 
 from astronomical objects provides a bound of  $\beta_{\gamma}>1.1\times10^{9}$  ~\cite{Bur08}.

The results we have presented here for solar chameleons are
only valid for values of the matter coupling below the resonance threshold
in the production mechanism at the tachocline ($\beta_{\rm m}<\nolinebreak10^{6}$).
For larger values of the matter coupling, the large values of the mass of
the chameleon inside the tachocline compared to the plasma mass lead
to a large suppression. The CHASE experiment is sensitive to the photon
coupling up to large values of the matter coupling ($\beta_{\rm m}\!\sim\!10^{14}$).
The region above $\beta_{\rm m}=1.9\times10^{7}$ is already excluded by the neutron experiments.
At low $\beta_{\rm m}$, our results extend the CHASE coverage by over three
orders of magnitude to below $\beta_{\rm m}=10$ into a region already
excluded by torsion pendulum bounds.

Higher values of $n$ could be envisaged but would not alter the physical picture discussed here
(see~\cite{Bra10} for a discussion of the $n=4$ case). Our results are to a large
extent insensitive to $n$  (Table~\ref{tab2}), provided we are only interested in the
region of parameter space below the resonance in the matter coupling.

\begin{table}[tbh]
\begin{centering}
\begin{tabular}{|c|c|c|}
\hline
index $n$ & $\beta_{\gamma}$ at $95\%$ CL\tabularnewline
\hline
\hline
1 & 9.26$\times$10$^{10}$\tabularnewline
\hline
2 & 9.21$\times$10$^{10}$\tabularnewline
\hline
4 & 9.20$\times$10$^{10}$\tabularnewline
\hline
6 & 9.19$\times$10$^{10}$\tabularnewline
\hline
\end{tabular}
\par\end{centering}
\begin{centering}
\protect\caption{Upper limit on $\beta_{\gamma}$ derived at CAST for different values of the index $n$ which defines the chameleon model. }
\par\label{tab2}\end{centering}
\end{table}

We studied the uncertainties in the assumptions for the solar model and their effect on the CAST result. 
If for example the solar luminosity bound is reduced by a factor 10, $\beta_{\gamma}^{\rm sun}$ is reduced by a 
factor 10$^{1/2}$, whilst $\beta_{\gamma}$ remains constant, resulting in a weaker limit relative to the solar luminosity bound.
Rather conservatively, the details of the radial field strength and its distribution at the tachocline may affect the
 $\beta_{\gamma}$ limit by a factor of $1.6$ (Table~\ref{tab1}). For the uncertainty on the magnitude 
 of the magnetic field at the tachocline we have considered a range from 4 to 25-30$\,$T, 
 which produces an uncertainty in $\beta_{\gamma}$ of a factor of about 1.6 up and down respectively (Fig.~\ref{Allbeta}).

All in all, we find that the chameleon parameter space has been significantly
reduced. Additional CAST data with the InGrid detector and an X-ray telescope will improve the photon
coupling sensitivity beyond the solar bound in the near future.
In parallel CAST is developing a detection technique which exploits the coupling of chameleons to matter.
 Chameleons of solar origin, focused by an X-ray telescope on CAST,
can be directly detected by a radiation pressure device~\cite{Bau14}.

\section{Conclusions}

CAST has made a first dedicated sub-keV search for solar chameleons based
on the Primakoff effect. This search, running in a vacuum configuration using a
readily-available apparatus, did not observe an excess above background and
has set a limit for the coupling strength to photons which for $n\ge1$ excludes a
new region of parameter space covering 3 orders of magnitude in matter coupling
and reaches down to the level of photon coupling corresponding to both the 10\% solar
luminosity bound and also the limit derived by CHASE.

\section{Acknowlegments}

We thank CERN for hosting the experiment and for the technical support to operate the magnet and the cryogenics.
We thank CERN PH-DT and TE-CRG groups for technical support to build the X-ray detector system and A. Niculae (PNDetector) for technical advice.

We acknowledge support from NSERC (Canada), MSES (Croatia), CEA (France), BMBF (Germany) under the grant numbers 05 CC2EEA/9 and 05CC1RD1/0 and DFG (Germany) under grant numbers HO 1400/7-1 and EXC-153, GSRT (Gree\-ce), NSRF: Heracleitus II, RFFR (Russia), the Spanish Ministry of Economy and Competitiveness (MINECO) under Grants No. FPA2008-03456, No. FPA2011-24058 and EIC-CERN-2011-0006. This work was partially funded by the European Regional Development Fund (ERDF/FEDER), the European Research Council (ERC) under grant ERC-2009-StG-240054 (T-REX), Turkish Atomic Energy Authority (TAEK), NASA under the grant number NAG5-10842. Part of this work was performed under the auspices of the U.S. Department of Energy by Lawrence Livermore National Laboratory under Contract No. DE-AC52-07NA27344. P.~Brax acknowledges partial support from the European Union FP7 ITN INVISIBLES (Marie Curie Actions, PITN- GA-2011- 289442) and from the Agence Nationale de la Recherche under contract ANR 2010 BLANC 0413 01.


\bibliography{Version30}

\end{document}